\documentclass[twocolumn]{aastex63}
\usepackage{amsmath}

\submitjournal{ApJ}
\accepted{2020/11/18}

\newcommand{\HI}{\ion{H}{1}~}
\newcommand{\kms}{km~s$^{-1}~$}
\newcommand{\Lya}{Ly$\alpha$ }

\newcommand{\Ha}{H$\alpha$ }

\shorttitle{Discovery of a low-z DLA using a starburst galaxy background source}
\shortauthors{Dupuis et al.}

\begin{document}

\title{Discovery of a Low-Redshift Damped \Lya System in a Foreground Extended Disk Using a Starburst Galaxy Background Illuminator}

\email{chris.dupuis@asu.edu}

\author[0000-0003-1739-3640]{Christopher M. Dupuis}
\affil{School of Earth and Space Exploration, Arizona State University, 781 Terrace Mall, Tempe, AZ 85287, USA}

\author[0000-0002-2724-8298]{Sanchayeeta Borthakur}
\affiliation{School of Earth and Space Exploration, Arizona State University, 781 Terrace Mall, Tempe, AZ 85287, USA}

\author[0000-0002-3472-0490]{Mansi Padave}
\affiliation{School of Earth and Space Exploration, Arizona State University, 781 Terrace Mall, Tempe, AZ 85287, USA}

\author[0000-0003-1268-5230]{Rolf A. Jansen}
\affiliation{School of Earth and Space Exploration, Arizona State University, 781 Terrace Mall, Tempe, AZ 85287, USA}

\author[0000-0003-2830-0913]{Rachael M. Alexandroff}
\affiliation{CITA; Dunlap Institute, University of Toronto, Toronto, ON, Canada}

\author[0000-0001-6670-6370]{Timothy M. Heckman}
\affiliation{Department of Physics \& Astronomy, Johns Hopkins University, Bloomberg Centre, 3400 N. Charles Street, Baltimore, MD 21218, USA}

\begin{abstract}
    We present the discovery of a low-redshift damped \Lya (DLA) system in the spectrum of background starburst galaxy SDSS~J111323.88+293039.3 ($z=0.17514$).
   The DLA is at an impact parameter of $\rm \rho=36~kpc$ from the star forming galaxy, SDSS J111324.08+293051.2 ($z=0.17077$). We measure an \HI column density of $N($\ion{H}{1}$)\rm =3.47\times10^{20}~cm^{-2}$ along with multiple low-ionization species such as \ion{N}{1}, \ion{N}{2}, \ion{Si}{2}, \ion{C}{2}, and \ion{Si}{3}. We also make an estimate of the covering fraction to be 0.883, giving us a limiting size of the DLA to be $A_{DLA}\rm \geq3.3~kpc^2$. Assuming a uniform column density over the entire DLA system, we estimate its mass to be $M_{DLA}\geq5.3\times 10^6~M_\odot$. 
   The extended illuminator and the low redshift of this DLA give us the unique opportunity to characterize its nature and the connection to its host galaxy.
   We measure a velocity offset of +131~km~s$^{-1}$ from the systemic velocity of the host for the DLA.
   This velocity is $-84$~km~s$^{-1}$ from the projected rotation velocity of the host galaxy as measured using a newly constructed rotation curve. Based on the size of the host galaxy, the \HI column density, and the gas kinematics, we believe this DLA is tracing the warm neutral gas in the \HI disk of the foreground galaxy.
   Our detection adds to a small set of low-redshift DLAs that have confirmed host galaxies, and is the first to be found using an extended background source.

\end{abstract}

\section{Introduction} \label{sec:intro}

Damped \Lya (DLA) systems are clouds of neutral hydrogen with column densities $N($\ion{H}{1}$)\geq2\times10^{20}\: \rm cm^{-2}$ \citep{wolfe86}. At these column densities, DLAs self-shield against photoionization from the cosmic ultraviolet background and therefore these clouds directly trace the neutral gas content in the universe \citep{wolfe05}. 
At higher redshifts, 
DLAs 
are believed to be an important tracer of gas that fuels star formation and drives galaxy growth \citep[e.g.][]{s-l00}. 
The damped \Lya feature is one of the strongest absorption features produced by parcels of intervening gas and therefore can be identified even in low signal-to-noise spectra. 
DLAs are a crucial tool for tracing the bulk of the cold gas content that cannot be imaged via the \HI 21cm hyperfine transition and their damping wings enable us to accurately estimate the gas column density. 
To trace gas inflows into galaxies, it is essential that surveys identifying DLAs are unbiased and not contingent upon metal-line selection criteria.
Such surveys are particularly important for tracing gas flows that are a major source of fuel for star formation and that may not yet be sufficiently enriched with metals \citep{Cooke11, Fum11, Raf12, Raf14, Sim12}. 

The majority of known DLAs were detected as absorption features in the spectra of background quasi-stellar objects (QSOs) at redshifts greater than 2. As cosmic expansion redshifts the \Lya transitions into the optical regime for $\rm z>2$, it is possible for ground-based spectroscopic surveys to detect them \citep[e.g.][]{croom04, hewett95}.
Of note is the Sloan Digital Sky Survey \citep[SDSS; ][]{york00}, that observed $\sim$300,000 QSOs and detected $\sim$27,000 DLAs \citep{proch05, paris17, mas17}. These efforts have dramatically increased the number of known DLAs and in particular at $\rm z\sim$ 2--4. 
However, there is a serious limitation to using QSOs as background sources. The space density of QSOs peaks at $\rm z\sim2.7$ resulting in fewer sightlines to search for DLAs at high redshifts \citep{schmidt95}. 
In contrast, star-forming galaxies are ubiquitous at high redshifts and recent studies have shown them to be useful background illuminators to probe diffuse gas structures \citep[e.g.][]{adelb05, steidel10, rubin10, bord11, cooke15, mawa16, lee16, stanic16, per18, rubin18a, rubin18b, chen20}. 
Star-forming galaxies release significant amounts of far-ultraviolet radiation ($\rm \lambda \ge 912~\AA$) that can trace neutral hydrogen (\Lya at $\rm 1215.67~\AA$) and other metal-line transitions within a considerable path-length. 
Star-forming galaxies may therefore be our best option for studying the gaseous media such as the circumgalactic medium (CGM) and the intergalactic medium (IGM) in many parts of the universe, including their own large-scale environments. 

Recent studies by \cite{cooke15} and \cite{mawa16} have detected high redshift DLAs ($\rm z\sim2.4$ and $\rm z\sim 3.3$, respectively) using background galaxy sightlines. Extended background sources like galaxies enable us to measure the sizes of absorbers using the residual flux technique and, when combined with the column density, provide the gas mass of the absorbing cloud. 
Previous studies have used multiple QSO sightlines to probe the extent of the DLAs \citep[e.g.][]{ellison07,monier09, cooke10, rubin15, krog18}, however QSO sightlines tend to be tens to hundreds of kpc apart in projection, thus making it difficult to probe the extent of any individual cloud. 
Galaxies, on the other hand, are extended and have sizes in the range of sub-kpc to kpc scales at the rest-frame of the absorber.
 \citeauthor{cooke15} and \citeauthor{mawa16} placed the first constraints on the size of DLAs detected in the spectra of background galaxies; $\rm \sim1-100~kpc^2$. 
 However these studies could not resolve the background source owing to its high-redshift and consequently had to assume a half-light radius based on statistics from galaxies with similar luminosities and morphologies. 
 Using DLAs found in the spectra of galaxies that are resolved at the same frequency band as the \Lya absorption-line will be a powerful technique for characterizing the sizes of absorbing clouds---something that cannot be done via QSOs as background illuminators.

Measuring the sizes of DLAs will be especially useful at low-redshifts where it is possible to image their surrounding environment in detail. However, very few DLAs have been detected at low-z primarily due to the expensive nature of blind UV surveys requiring space based observatories. These few known low-z DLAs have played an important role in our understanding of how these systems are connected to galaxies. 
Low-z DLAs have been found (a) in group environments, where they may be best explained by interactions between group members forming tidal streams and bridges \citep[e.g.][]{kacp10, augustin18, chen19}, (b) in areas with no obvious host galaxies, where they may be due to optically faint, gas-rich, dwarfs \citep[e.g.][]{battist12, kanek18}, or (c) DLAs found to be associated with galactic disks \citep[e.g.][]{gupta13, gupta18}.

Here we present an analysis of the
first low-redshift DLA detected in the spectrum of a background star-forming galaxy \citep{alex15}, the starburst galaxy SDSS J111323.88+293039.3 ($z_{\rm BG}=0.17514$; hereafter BG). The BG sightline passes a projected distance of 36~kpc from the DLA host galaxy, the foreground $L_*$ galaxy SDSS J111324.08+293051.2 ($z_{\rm FG}=0.17077$; hereafter FG). This discovery is an extremely special case for three reasons. 
First, it adds to our limited sample of low-redshift DLA absorbers whose environments can be imaged in detail. This is critical for investigating the connection between the DLAs and their host galaxies and can only be achieved for low-redshift systems due to surface brightness dimming with redshift. Second, the background galaxy is resolved at rest-frame ultraviolet wavelengths at the redshift of the DLA, thus enabling us to estimate the size and mass of the DLA system. Third, this discovery also marks the successful demonstration of the technique of using background galaxies to probe low-z DLAs. Since at these redshifts DLAs are not expected to be pristine, such systems could be easily targeted via optical absorption features such as \ion{Na}{1}~D and \ion{Ca}{2} features in the era of large optical observatories like the Giant Magellan Telescope (GMT), the Thirty Meter Telescope (TMT), and the Extremely Large Telescope (ELT).

The remainder of the paper is organized into three sections. We present our ground- and space-based observations and data analysis in Section 2, where we also discuss the background source. This is followed by a discussion on the nature and origin of the DLA in Section \ref{sec:discussion}. We summarize our conclusions in Section \ref{sec:conclusion}. All the values presented in this study have been calculated using the following cosmological parameters: $H_0 =70~{\rm km~s}^{-1}~{\rm Mpc}^{-1}$, $\Omega_m = 0.3$, and $\Omega_{\Lambda} = 0.7$. 

\section{Target, Observations, and Data Analysis} \label{sec:data}

\begin{figure*}[!t]
    \centering
    \includegraphics[width=\linewidth]{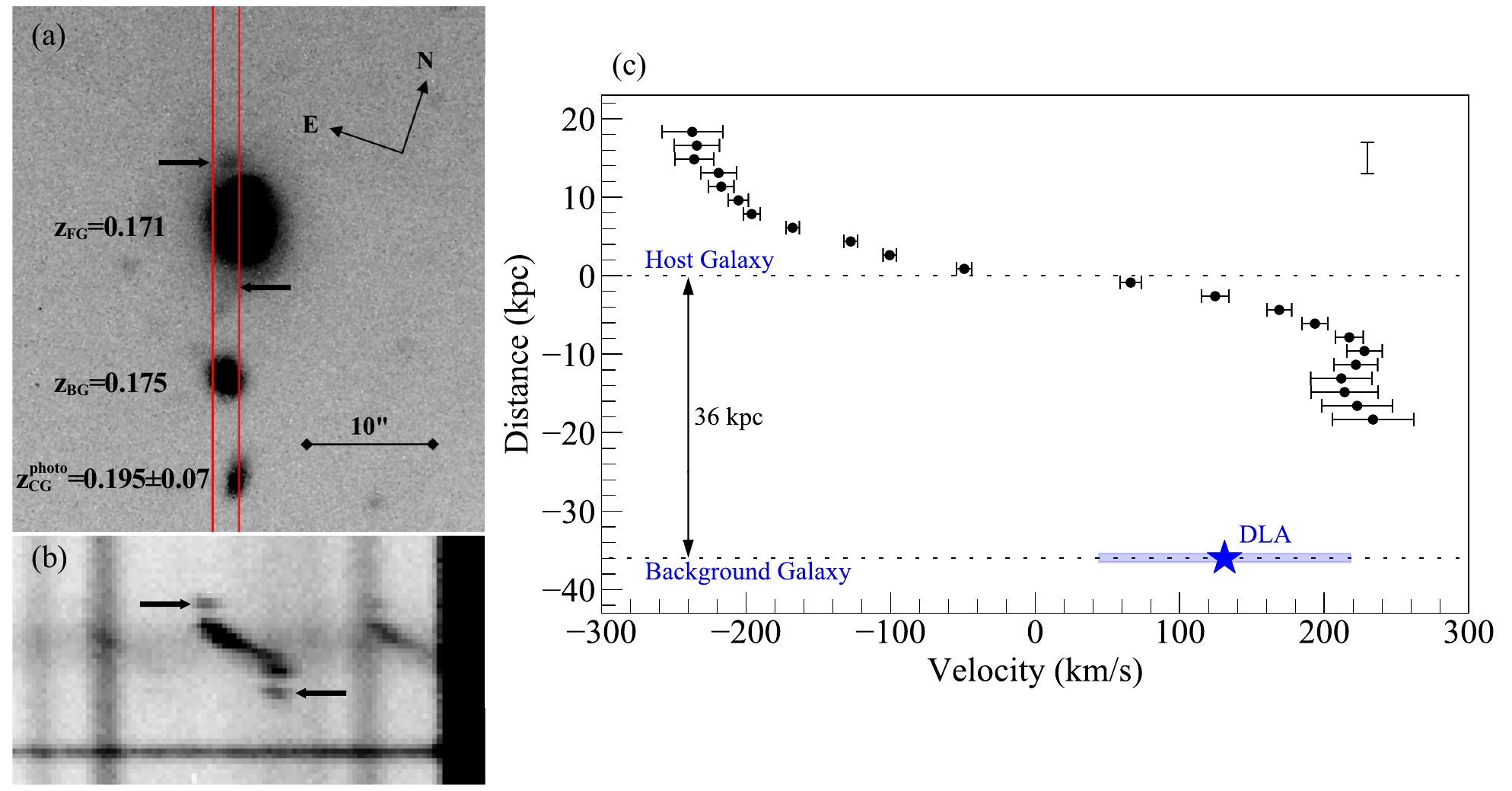}
    \caption{(a) MMTCam \textit{r}-band image of the J1113+2930 galaxy field. The $180\arcsec \times 2\arcsec$ slit used in our Blue Channel Spectrograph observations is overlayed in red. The slit was centered on BG and was oriented towards FG to cover the extension seen to the south. (b) Two-dimensional spectrum of FG and BG cropped around the H$\alpha$ emission of FG showing a clear signature of rotation. The emission is resolved over the full extent of the galaxy and extends more than 12\arcsec on the sky. Multiple sky lines can also be seen as well as signatures of the [\ion{N}{2}] doublet on either side of the H$\alpha$ emission from FG. There are also two distinct areas of emission marked with black arrows that are separate from the main galaxy. Their symmetry suggests a possible ring of star formation around FG. The approximate location of this emission is marked by arrows in the $\textit{r-}$ band image (Figure \ref{fig:rotation}a). (c) Galaxy rotation curve of FG constructed using the H$\alpha$ emission line. Assuming the galaxy is rotationally symmetric, we calculate a new redshift for the galaxy of $z_{\rm FG}=0.17077 \pm 0.00018$. The relative velocity of the dominant DLA component with respect to the rest-frame of FG is shown by the blue star with the shaded regions representing 90\% of the width of the \ion{Si}{2} $\lambda1260$ profile. The DLA is 18 kpc from the edge of the \Ha disk at a velocity of 131~km~s$^{-1}$.}
    \label{fig:rotation}
\end{figure*}

\subsection{Ground-Based Observations}
\subsubsection{Imaging}
 We imaged the DLA field with the 1.8-m Vatican Advanced Technology Telescope (VATT) on UT 2019 March 31 in SDSS \emph{ugri} with exposure times of 2$\times$900\,s, 2$\times$600\,s, 2$\times$600\,s, and 2$\times$600\,s, respectively. 
 The Vatt4k Imager has a $4064\times 4064$ pixel back-illuminated STA0500 CCD that is binned 2$\times$2 on read-out, giving a pixel scale of 0$\farcs$375 and a field of view (FOV) of $\sim$12$\farcm$5 on a side.
 For calibrations both dome and sky flats were taken in the \textit{g-} and \textit{r-}bands while only sky flats were taken in the \textit{u-} and \textit{i-}bands. Due to time constraints only three twilight flats were taken in the \textit{u-} and \textit{i-}bands. The average seeing for these observations were $\rm FWHM\sim1\farcs00$.

 Higher resolution imaging was obtained with the 6.5-m MMT Observatory (MMTO) using the MMTCam on UT 2020 February 17 with images acquired in SDSS \emph{ugr}. Due to persistent thin clouds we were unable to resolve the galaxies in the \textit{u-}band. Exposure times ranged from $\rm180\,s~(\emph{gr})$ to $\rm 600\,s~(\emph{u})$ and seeing conditions were $\rm FWHM\sim0\farcs8$. Twilight sky flats were taken in all filters for calibrations. Dark frames were obtained at exposure times corresponding to the science images due to a non-negligible dark current. The MMTCam has a FOV of $\sim$2$\farcm$7 on a side and contains a back-illuminated CCD that is $2048\times 2048$ pixels. The detector was binned 2$\times$2 on read-out, giving a pixel scale of 0$\farcs$16.
 
 Data reduction for both data sets were completed using standard IRAF procedures. Science frames were bias subtracted and flat fielded using sky and dome flats (where available). MMTCam science images were additionally dark current subtracted. Cosmic rays were removed using \texttt{L.A. Cosmic} \citep{lacosmic}. Science images were shifted and aligned using the \texttt{imregister} routine \footnote{\url{http://www.public.asu.edu/\~rjansen/iraf/rjtools.html}} allowing for stacks to be created in each filter. The reduced and stacked \textit{r-}band image obtained with MMTO can be seen in Figure \ref{fig:rotation}a.

 \subsubsection{Spectroscopy and FG Rotation Curve}\label{subsec:spec}

To measure the rotation of the DLA host galaxy, FG, we acquired long-slit spectra in 2020 February using the Blue Channel Spectrograph at the MMTO. We utilized the 832 lines mm$^{-1}$ grating in 1st order centered on the redshifted \Ha line of FG (7682 \AA) using the R-63 blue blocking filter. The wavelength coverage of the spectra were 6734--8642~$\rm \AA$. All spectra were obtained with the $180\arcsec \times 2\arcsec$ slit with the data binned $2\times1$ resulting in a spatial pixel scale of 0$\farcs$6, a dispersion scale of 0.72 $\rm \AA$ pixel$^{-1}$, and a spectral resolution of $\rm \sim3.3~\AA$ (FWHM; $\sim$$\rm 150~km~s^{-1}$). The slit was centered on BG and the position angle was set to 18$^\circ$ in order to cover the stellar extension seen in the south of FG as well as the galaxy itself (Figure \ref{fig:rotation}a). We obtained three spectra of the field with exposure times of 1200~s.

The data were reduced using standard IRAF procedures. Flat fielding of the science frames was accomplished using internal quartz lamp illumination images. The sky was subtracted from the science frames during extraction of the galaxy spectra by fitting a polynomial to the median sky values in areas adjacent to the extraction region. The bright spectrum of BG was used as the trace during the extraction of the FG spectra. Wavelength calibration was completed using HeNeAr arc line lamp spectra that bracketed the science frames. The reduced two-dimensional spectrum of FG and BG is shown in Figure \ref{fig:rotation}b. 

The rotation curve was constructed following the procedures of \cite{vogt96}, \cite{steidel02}, and \cite{kacp10mg2}. Individual one-dimensional spectra were extracted using apertures that were summed over three spatial columns ($\sim1\farcs8$) and then incremented over one spatial pixel to cover the full length of the galaxy. Wavelength calibrations were made using HeNeAr arc line lamp spectra extracted over the same spatial pixels resulting in solutions accurate to $\sim$0.1~$\rm \AA$. A single Gaussian was fit to each \Ha emission line to determine the wavelength centroid for each spatial column. The rms uncertainties of the Gaussian fit were used to estimate the uncertainties of the central wavelength and ranged from $\sim\rm 5-30~km~s^{-1}$. All spectra were vacuum and barycentric velocity corrected using the IRAF package \texttt{RVSA}. Figure \ref{fig:rotation}c shows the constructed FG rotation curve. 

We additionally extracted an integrated one-dimensional spectrum for FG to determine the galaxy redshift. This resulted in a value consistent with the SDSS determined value of $z_{\rm FG,SDSS}=0.17048$. However, plotting the rotation curve at the rest-frame of FG using this redshift resulted in the center of rotation being offset from $v=0$ by $\sim$87~km~s$^{-1}$. Due to this discrepancy we calculated a new redshift for FG of $z_{\rm FG}=0.17077 \pm 0.00018$ using the rotation curve assuming the galaxy is rotationally symmetric. 
A non-uniform distribution of \ion{H}{2} regions orbiting within FGs disk is the likely reason reason the rotation center is offset from the peak of the integrated \Ha spectrum.

We do not detect any emission (continuum/line) from the third galaxy in the field to the southwest of FG and BG which we label CG. Based on the orientation of the slit we should be partially covering CG (Figure \ref{fig:rotation}a). CG has an SDSS photometric redshift of $z_{\mathrm{phot}}=0.19\pm0.07$, so we would expect to see some emission if the slit was on the galaxy. However, as we cannot be certain of the exact location of the slit we refrain from making any estimates of the spectroscopic redshift.

\subsection{Background Source}

\begin{figure*}[t]
    \centering
    \includegraphics[width=\linewidth]{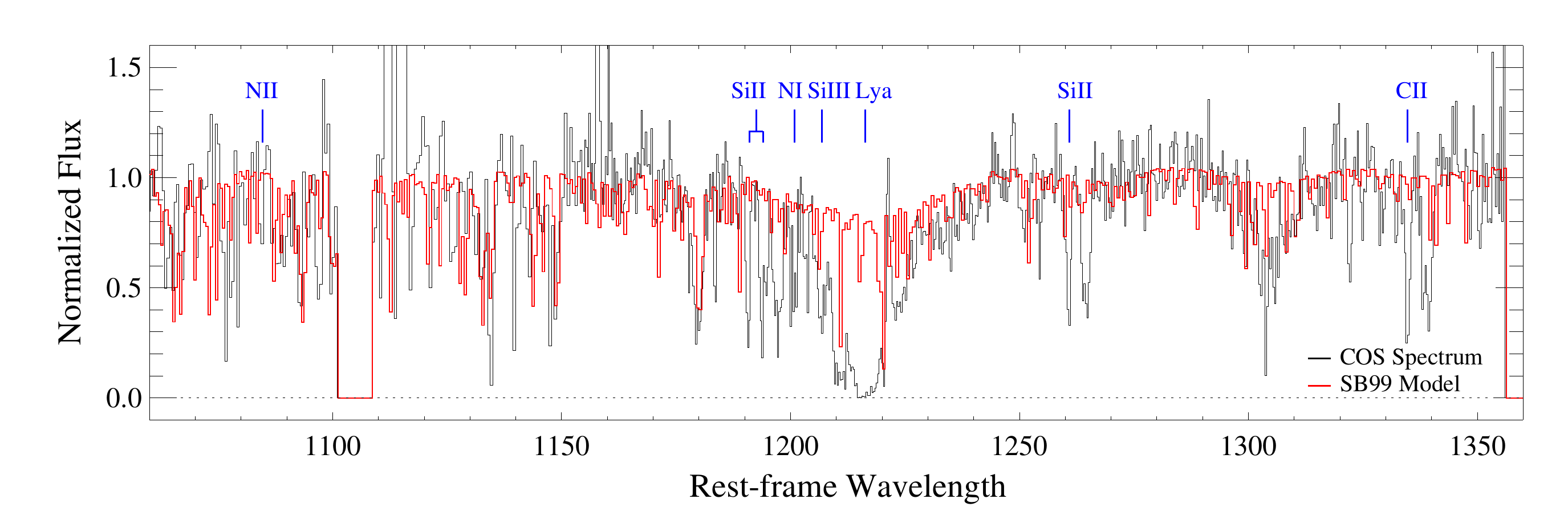}
    \caption{COS spectrum of the background galaxy (BG, black) plotted in the rest-frame of the foreground DLA host galaxy (FG, $z_{\rm FG}=0.17077$) with the best fit Starburst99 model published by \cite{alex15} shown in red. The spectrum is scaled to only show regions containing DLA absorption features. 
    The model shown represents an instantaneous starburst, a Kroupa IMF, and a young stellar population age of 22 Myr with solar metallicity. The detected DLA absorption features of \ion{N}{2} $\lambda1084$; \ion{N}{1} $\lambda\lambda$ 1199.5, 1200.2, 1200.7;} \ion{Si}{2} $\lambda\lambda$ 1190, 1193; \ion{Si}{3} $\lambda1206$; \ion{H}{1} $\lambda1216$ (\Lya); \ion{C}{2} $\lambda1334$; and \ion{Si}{2} $\lambda 1260$ are labeled in blue. To aid in clarity the spectrum shown here is binned by 15 pixels.
    \label{fig:sb99}
\end{figure*}

The DLA was found in the spectrum of BG and was first reported by \cite{alex15} as part of a study to characterize the properties of Lyman-break analog (LBA) galaxies using the Cosmic Origins Spectrograph (COS) aboard the \textit{Hubble Space Telescope (HST)} \citep[]{over09}. FG was identified as the host galaxy of the DLA based on its SDSS measured redshift and the location of the \Lya absorption profile. Figure~\ref{fig:sb99} shows the COS data in black and the best-fit stellar population model for BG generated by \citeauthor{alex15} using Starburst99 \citep{SB99,sb992} in red.
\citeauthor{alex15} used the \ion{C}{3} $\lambda1175$ absorption feature, that directly traces the stellar photosphere, to estimate the young stellar populations to be 22~Myrs old, assuming an instantaneous starburst, and a Kroupa IMF at solar metallicity. 

Using the model of FG allowed us to identify absorption features that were not consistent with stellar features; these are marked in blue in Figure \ref{fig:sb99}. 
This was an important step to avoid confusion in identifying intervening absorption from intrinsic features of BG. 
However, we did not subtract the stellar population model for fitting the \Lya feature of the DLA as it is close to the \Lya feature of BG and the Starburst99 modeling at these wavelengths are highly uncertain. This is primarily because the \Lya transition is a resonance line and the contribution of the interstellar medium (ISM) in changing the intrinsic shape of the profile cannot be fully modeled.
Fortunately, the damped \Lya feature is extremely broad and covers line-free regions. These regions along with constraining our profile fitting algorithm to obey symmetry allowed us to estimate the properties of the \Lya feature while removing the BG stellar features.

\subsection{COS Spectra and Line Fitting} \label{sec:fit}

The Ultraviolet spectra (Figure \ref{fig:sb99}) of BG were obtained using the G130M and G160M medium resolution gratings of COS (see \citet{alex15}). 
After processing with the standard COS pipeline, the G130M and G160M spectra were combined resulting in full coverage in observed wavelength from $\lambda=$ 1160--1790 \AA. 
We binned the spectrum by 7 pixels (the resolution of the data) resulting in a spectral bin size of $\sim$17\:km\:s$^{-1}$.
Once the spectrum was generated, visual inspection showed a damped \Lya absorption feature centered at 1424\:\AA.

 At the location of the DLA the spectrum covers $\rm \lambda_{rest}=$ 968--1524\:\AA. We identified all absorption features associated with either BG, the DLA, or the Milky Way's ISM through visual inspection.
 We determined the continuum bracketing the absorption features using absorption-free regions within $\pm$2000\:km\:s$^{-1}$ of the feature with the exception of the \Lya transition where the region was selected to be $\pm$8000\:km\:s$^{-1}$ of the systemic velocity due to the large width of the profile. 
 The choices for continuum regions were motivated by feature-free zones in the best fit Starburst99 model of the BG spectrum shown in Figure \ref{fig:sb99}. The continuum near each absorption feature was estimated using a Legendre polynomial between order 1 and 5 using a procedure similar to \cite{sem04} which was then used to produce the normalized spectrum.

\begin{figure*}[t!]
    \epsscale{1.2}
    \plotone{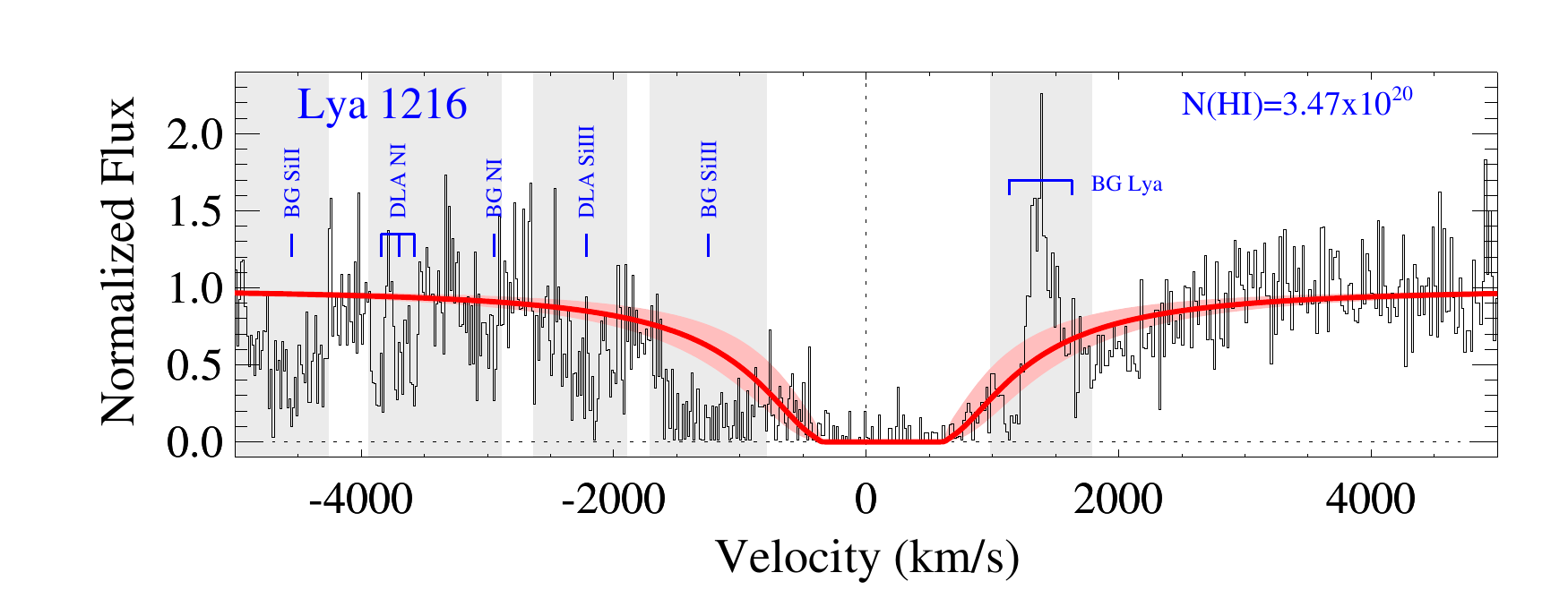}
    \caption{Damped \Lya absorption feature found in the spectrum of the background galaxy (BG; $z_{\rm BG}=0.17514$) centered at the location of the DLA host galaxy (FG; $z_{\rm FG}=0.17077$). The absorption is 131~km~s$^{-1}$ offset from the systemic of FG. The best fit Voigt profile is shown in red with the uncertainty in the column density shown by the pale red shaded regions. The inferred column density is shown in the top right corner. Contaminating absorption lines are marked in blue. These regions were masked from the Voigt profile fit (grey shaded regions). Additionally, due to the large amount of contamination the velocity centroid was fixed to the centroid of the strongest \ion{Si}{2} component to aid in fitting.}
    \label{fig:dla}
\end{figure*}

\begin{deluxetable*}{lccccc}
	\tablecaption{Transitions Associated with the DLA \label{tab:trans}}
	\tablecolumns{6}
	\tablehead{
	\colhead{Transition} & \colhead{$\rm \lambda_{rest}$} & \colhead{$\rm W_{rest}$\tablenotemark{a}} & \colhead{Centroid\tablenotemark{b}} & \colhead{Doppler $b$} & \colhead{log $N$ } \\
	\colhead{}	 & \colhead{(\AA)} & \colhead{(m\AA)} & \colhead{(km\:s$^{-1}$)} & \colhead{(km\:s$^{-1}$)} & \colhead{(log\:cm$^{-2}$)}
	}
	\startdata
	\ion{H}{1} & 1215.67 & 18538\tablenotemark{c} & 131\tablenotemark{d} & $130^{+144}_{-69}$ & $20.5^{+0.2}_{-0.2}$\tablenotemark{e} \\
	\ion{N}{1} & 1199.55 & $267\pm142$ & $125\pm14$ & $40^{+27}_{-16}$ & $\geq$14.5 \\
 	 & 1200.22 & $229\pm140$ \\
 	 & 1200.71 & $195\pm106$ \\
 	\ion{N}{2} & 1083.99 & $410\pm105$ & $-9\pm28$, $124\pm15$ & $43^{+118}_{-36}$, $45^{+31}_{-19}$ & $\geq14.1$, $\textgreater$14.6 \\
	\ion{Si}{2} & 1190.42 & $665\pm168$ & $21\pm16$, $131\pm11$ & $33^{+36}_{-18}$, $41^{+18}_{-13}$ & $\geq$13.5, \textgreater14.2 \\
 	& 1193.29 & $662\pm177$ \\
 	& 1260.42 & $596\pm135$ \\
 	\ion{C}{2} & 1334.53 & $740\pm195$ & $-3\pm25$, $113\pm26$ & $41^{+72}_{-19}$, $54^{+87}_{-34}$ & $\geq$14.3, \textgreater14.5 \\
 	\ion{Si}{3} & 1206.50 & $921\pm250$ & $-3\pm33$, $132\pm20$ & $65^{+130}_{-37}$, $40^{+98}_{-29}$ & $\geq$13.3, \textgreater13.4 \\
	\ion{Si}{4} & 1393.76 & \textless140\tablenotemark{f} & \ldots & \ldots & \textless14.1\tablenotemark{f} \\
	\ion{N}{5} & 1238.82 & \textless150\tablenotemark{f} & \ldots & \ldots & \textless14.2\tablenotemark{f} \\
	\ion{Ca}{2}\tablenotemark{g} & 3933.66 & \textless343\tablenotemark{f} & \ldots & \ldots & \textless13.1\tablenotemark{f} \\
	& 3968.47 & \textless220\tablenotemark{f} & & & \\
	\ion{Na}{1}\tablenotemark{g} & 5889.95 & \textless219\tablenotemark{f} & \ldots & \ldots & \textless12.6\tablenotemark{f} \\
	& 5895.92 & \textless185\tablenotemark{f} & & & \\
	\enddata
	\tablenotetext{a}{Equivalent widths as estimated directly from data. The error is calculated from both continuum fitting and statistical errors.}
	\tablenotetext{b}{Velocity centroid values are reported with respect to the rest-frame of FG ($\rm z=0.17077$).}
	\tablenotetext{c}{\ion{H}{1} equivalent width taken from Voigt profile fit.}
	\tablenotetext{d}{Due to contamination in the wings, \ion{H}{1} centroid fits resulted in values $\geq$150\:km\:s$^{-1}$ offset from the dominant feature in the metal line transitions. Due to these unrealistic values we fixed the \ion{H}{1} centroid to match the dominant \ion{Si}{2} absorption feature.}
	\tablenotetext{e}{\Lya column density error estimates are based on the minimum and maximum fits to the profile using different masking regions for the BG \Lya emission. See the discussion in Section \ref{sec:fit}.}
	\tablenotetext{f}{Upper limits on the equivalent widths were estimated from 3$\sigma$ uncertainties measured over 200~\kms. Column density upper limits were estimated from the limiting equivalent width assuming the linear regime of the curve of growth \citep{draine11}.}
	\tablenotetext{g}{\ion{Ca}{2} and \ion{Na}{1} upper limits are derived from SDSS optical spectra.}
\end{deluxetable*}

\begin{figure*}[th!]
    \centering
    \includegraphics[width=\linewidth]{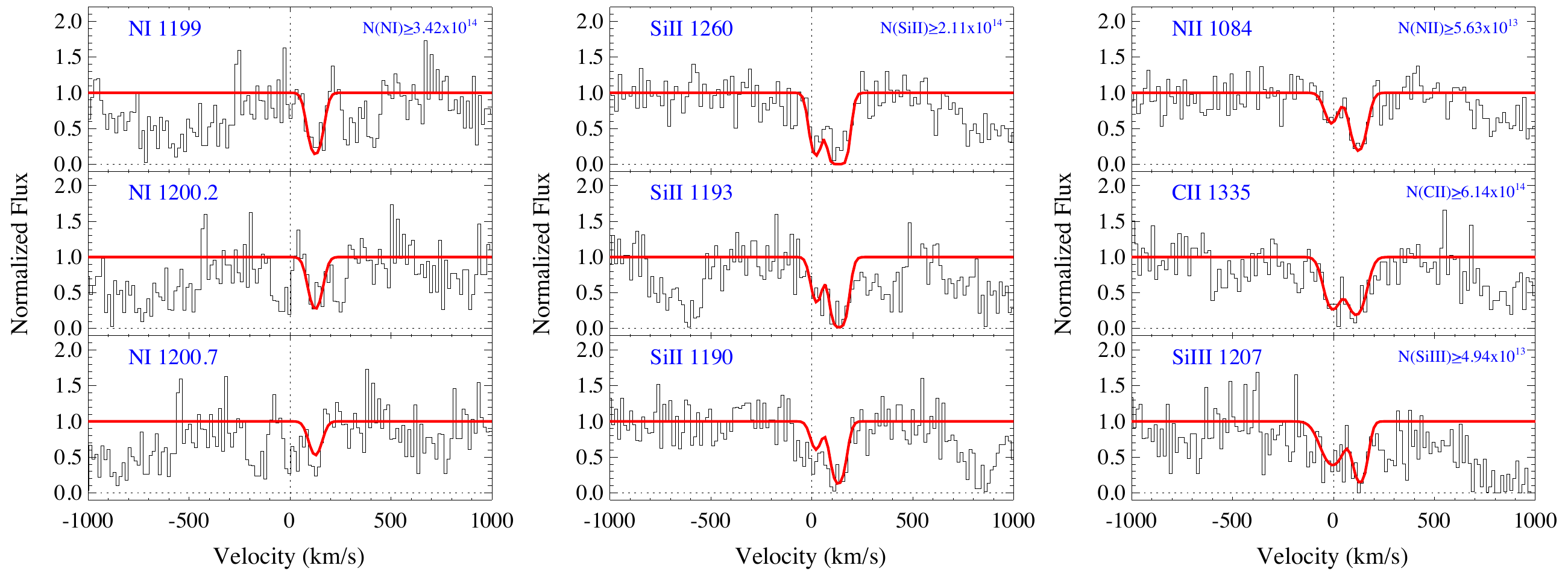}
    \caption{COS spectra of the metal-line transitions associated with the DLA plotted at the rest-frame of FG ($z_{\rm FG}=0.17077$). Best fit Voigt profiles for the transitions are plotted in red. We fit the three \ion{Si}{2} transitions simultaneously. Each transition has one component that is optically thick resulting in only lower limit estimates for the column densities that are marked in the upper right for each species.}
    \label{fig:metals}
\end{figure*}

We detected the following absorption lines associated with the DLA: \ion{H}{1} $\lambda1216$ (Ly$\alpha$);
\ion{N}{1} $\lambda\lambda$ 1199.5, 1200.2, 1200.7; \ion{N}{2} $\lambda1084$; \ion{Si}{2} $\lambda\lambda$ 1190, 1193, 1260; \ion{C}{2} $\lambda1335$; and \ion{Si}{3} $\lambda1207$. Figures \ref{fig:dla} and \ref{fig:metals} show the \Lya and metal-line absorption features of the DLA. We fit Voigt profiles to each feature using the software of \cite{fitz97} and techniques similar to \cite{trip08} and \cite{tum13}. The fits derived the velocity centroids, Doppler \textit{b}-values, and column densities for each transition, as listed in Table \ref{tab:trans}. We fit two components for all metal-line transition, with the exception of \ion{N}{1}, as visual inspection showed there were two features in each profile. For \ion{N}{1} and \ion{Si}{2} the optimal fit was calculated simultaneously for the three transitions in each species by using their relative oscillator strengths. This was especially critical for \ion{Si}{2} $\lambda1193$ as it was blended with the \ion{Si}{2} $\lambda1190$ absorption feature of BG. The measurement uncertainties were derived using the error analysis methods of \cite{sem92}.

The \Lya feature of the DLA is blended with \Lya emission from BG at $\sim$1200\:km\:s$^{-1}$ as well as with \ion{N}{1}, \ion{Si}{2}, and \ion{Si}{3} absorption from both the DLA and BG between $-$5000 and $-$100\:km\:s$^{-1}$ (see Figure \ref{fig:dla}) resulting in severe contamination of the wings of the \Lya absorption profile. 
We fit the Voigt profile by first masking the contaminated regions of the spectrum (grey regions in Figure \ref{fig:dla}), and then manually fixing the centroid to the dominant component derived from the low-ionization metal-line transition \ion{Si}{2}. The ionization potential of \ion{Si}{2} is 16.3~eV and is the closest to the ionization potential of \HI among all the other metal-line species that we detected in this system. In addition, the estimate of the centroid for \ion{Si}{2} is most reliable as it was derived by fitting three transitions of varying optical depth simultaneously unlike single transition fits for \ion{Si}{3} and \ion{C}{2}. As a majority of the blueward wing is contaminated by \ion{Si}{3} absorption, the red wing was the main driver for the shape of the profile. The choice of masking regions on this wing therefore had a large effect on the resulting best fit. This is especially true near 1100\:km\:s$^{-1}$ where the increased photon count could be following the absorption profile or be due to blue-shifted \Lya emission from BG that is seen in other LBA galaxies from the \citeauthor{alex15} study. However, we cannot know the true extent of the blending here, and have therefore gauged the uncertainty in the inferred \HI column density from a series of fits with varying, but reasonable, choices for the wavelength limits of the masking regions.
 
The best fit Voigt profile gives a column density of \HI of $\mathrm{log}N($\ion{H}{1}$)=20.5^{+0.2}_{-0.2}$, confirming the cloud as a DLA. The \ion{H}{1} column density is solely based on the \Lya transition as Ly$\beta$ is blended with emission due to the Milky Way. The three transitions each of \ion{N}{1} and \ion{Si}{2}, as well as \ion{N}{2} and \ion{C}{2} are the other low-ionization species detected in this system. We find a single component for \ion{N}{1} at 125\:km\:s$^{-1}$ while we detect two components for the other transitions: at $-$9 and 124\:km\:s$^{-1}$ for \ion{N}{2}, 21 and 131\:km\:s$^{-1}$ for \ion{Si}{2}, and at $-$3 and 113\:km\:s$^{-1}$ for \ion{C}{2}. Most of the absorption shows strong signs of saturation leading to lower limits in their estimated total column densities of $\mathrm{log}N($\ion{N}{1}$)\geq14.5$, $\mathrm{log}N($\ion{N}{2}$)\geq14.7$, $\mathrm{log}N($\ion{Si}{2}$)\geq14.2$ and $\mathrm{log}N($\ion{C}{2}$)\geq14.7$. The intermediate-ionization transition \ion{Si}{3} also shows two components at -3 and 132\:km\:s$^{-1}$ offset from FG with a column density of $\mathrm{log}N($\ion{S}{3}$)\geq13.6$. Interestingly, the relative strength of the two components varies between the transitions. 
The ratio changes from 1:5 for \ion{Si}{2} with an ionization potential of 16.3~eV, 1:3 for \ion{N}{2} (29.6~eV), 2:3 for \ion{C}{2} (24.4~eV), to 4:5 for \ion{Si}{3} (33.5~eV). \ion{C}{2} likely suffers from saturation in the blue component and as a result does not follow the general trend. However, the trend of these ratios combined with no detectable blue component for \ion{N}{1} (ionization potential of 14.5 eV), indicates that the blue component is at a higher ionization state than the red component. Since most of the absorption features are saturated (optically thick), a detailed modeling of the ionization state of the gas is not possible. 

Our data do not show absorption at 3$\sigma$ or higher significance for \ion{N}{5} $\lambda1239$ or \ion{Si}{4} $\lambda1394$ resulting in upper limits of $\mathrm{log}N($\ion{N}{5}$)\leq14.2$ and $\mathrm{log}N($\ion{Si}{4}$)\leq14.1$. No estimate for \ion{O}{1} $\lambda1302$ could be made due to contamination. Additionally, we analyzed the SDSS spectrum of BG and do not detect features corresponding to \ion{Ca}{2} $\lambda\lambda3934$, 3968 or \ion{Na}{1} $\lambda\lambda5890$, 5896. After fitting the continuum in the same method as described above, we estimate the limiting column densities as $\mathrm{log}N($\ion{Na}{1}$)\leq 12.6$, and $\mathrm{log}N($\ion{Ca}{2}$)\leq 13.1$.

\section{Results and Discussion} \label{sec:discussion}

\subsection{Galaxy Properties} \label{sec:system}

\begin{figure}
    \centering
    \includegraphics[width=\linewidth]{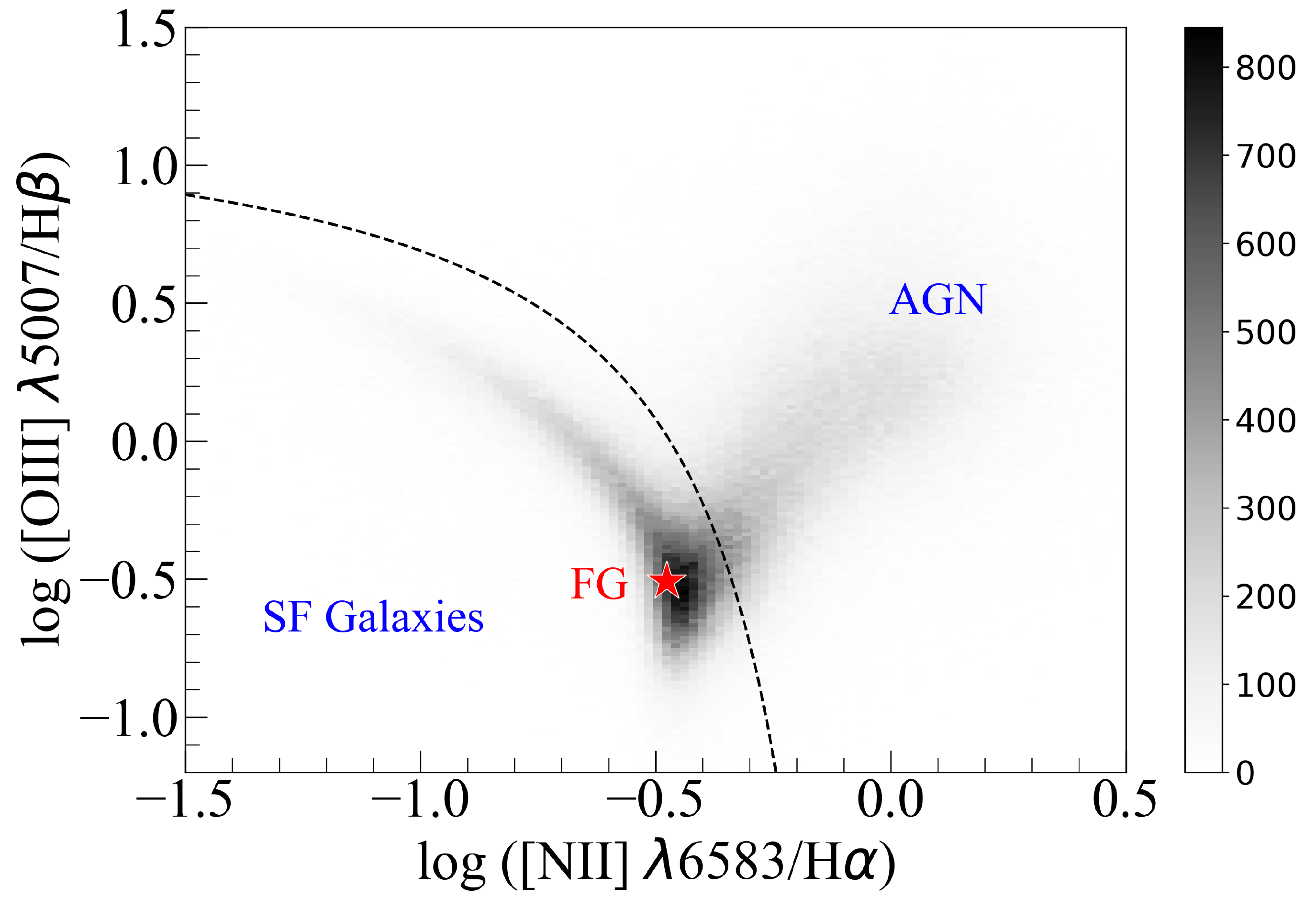}
    \caption{Location of the DLA host galaxy (FG, red star) on the BPT diagram \citep{bpt81}. Also plotted is a 2D-histogram of SDSS galaxies (grey). The flux data are from the MPA-JHU DR7 catalog. The dashed curve marks the demarcation between star-forming galaxies and AGN as defined by \cite{kauf03}. From the location of FG it is clear the line emission is powered by star formation.}
    \label{fig:BPT}
\end{figure}

 More detailed properties of the three galaxies in the field, BG, FG, and CG are listed in Tables \ref{tab:galaxies} and \ref{tab:phot}. The derived quantities including the half-light radii and FG inclination were estimated using our MMT $r$-band image with GALFIT (\citep{galfit1, galfit2}. The DLA host galaxy, FG, is a 3.8~L$_{\star}$ galaxy with a stellar mass of $1.26\times10^{11}~M_\odot$. We estimate the halo mass of the galaxy to be $3.98\times10^{12}~M_\odot$ with a virial radius of 361~kpc using the prescriptions of \citet{kravtsov13, liang2014, mandelbaum16}. The BG sightline passes FG at an impact parameter of $\rho=36$\:kpc implying that the sightline is probing an extended disk or inner CGM of FG at 0.1~$R\rm _{vir}$. Additionally, the position angle of FG of 27.2$^\circ$ implies the BG sightline is probing FG within a projected $\sim$14$^\circ$ from the major axis of the galaxy. FG is highly star forming with a specific star formation rate of log~sSFR$=-9.85$\:log\:yr$^{-1}$. We conclude that FG does not have an active galactic nucleus (AGN) based on emission line ratios that unambiguously places it in the region occupied by star forming galaxies on 
an BPT emission line ratio diagnostic
diagram (see Figure \ref{fig:BPT}) \citep{bpt81}. The emission-line fluxes were adopted from the MPA-JHU value-added catalog\footnote{\url{https://wwwmpa.mpa-garching.mpg.de/SDSS/DR7/}} that was generated using SDSS DR7 (see Table \ref{tab:phot}). The dotted curve is the demarcation between star forming galaxies and AGN defined in \cite{kauf03}.

A third galaxy, SDSS J111323.63+293032.0 (CG), is seen to the southwest of BG and FG (see Figure \ref{fig:rotation}a). No spectroscopic redshift is available for CG, but its SDSS photometric redshift of $z_{\mathrm{phot}}=0.19\pm0.07$ places it near FG within the error. This galaxy is a star-forming galaxy with an \textit{r-}band magnitude of 20.22 and a $g-r$ color of 0.51. The projected separations of CG at the location of the DLA from FG and BG are 66~kpc and 30~kpc, respectively. Since we do not have spectroscopic confirmation of the redshift, we refrain from assigning CG as the host of the DLA, although we do consider the possibility of the DLA originating in a tidal structure between FG and CG (see \S~\ref{sec:origin}).

 \begin{deluxetable*}{lccccccc}
	\tablecolumns{7}
    \tablecaption{Properties of the Background Source and DLA Host \label{tab:galaxies}}
    \tablehead{
    \colhead{Label} & 
    \colhead{Galaxy} & 
    \colhead{R.A.} & 
    \colhead{Decl.} & 
    \colhead{$z$} & 
    \colhead{log M$^*$\tablenotemark{a}} &
    \colhead{log sSFR\tablenotemark{a}} & \colhead{SFR$_{\rm FUV}$\tablenotemark{b}} \\
    \colhead{} & \colhead{} & \colhead{} & \colhead{} & \colhead{} & 
    \colhead{(log~$M_{\odot}$)}  &  
    \colhead{($\rm log~yr^{-1}$)} & \colhead{($M_{\odot}$~$\rm yr^{-1}$)}
    }
	\startdata
	BG & J111323.88+293039.3 & 168.3495 & 29.5109 & 0.17514 & 9.6 & -9.15 & 7.09 \\
	FG & J111324.08+293051.2 & 168.3503 & 29.5142 & 0.17077\tablenotemark{c} & 11.1 & -9.85 & \ldots  \\
	CG & J111323.63+293032.0 & 168.3484 & 29.5089 & $0.195\pm0.0652$\tablenotemark{d} & \ldots & \ldots & \ldots
	\enddata
    \tablenotetext{a}{M$^*$ and sSFR values are taken from the MPA-JHU DR7 catalog. Values are corrected for fiber placement.}
    \tablenotetext{b}{Star formation rate using FUV luminosity corrected by far-Infrared luminosity, calculated in \cite{alex15}.}
    \tablenotetext{c}{FG redshift calculated using our newly constructed galaxy rotation curve. See Section \ref{subsec:spec} and Figure \ref{fig:rotation}c.}
    \tablenotemark{d}{Photometric redshift of CG from SDSS.}
\end{deluxetable*}

\begin{deluxetable*}{lcccccccccc}
	\tablecolumns{7}
    \tablecaption{Photometry, Line Fluxes, and Derived Quantities} \label{tab:phot}
    \tablehead{
    \colhead{Label} &
    \colhead{$g-r$} &
    \colhead{$r$} &
    \colhead{$R_{50,r}$\tablenotemark{a}} &
    \colhead{$R_{\rm 50,FUV}$\tablenotemark{b}} &
    \colhead{$i$\tablenotemark{a}} &
    \colhead{$v_{\rm circ}$} &
    \colhead{$f_{\rm H\alpha}$\tablenotemark{c}} &
    \colhead{$f_{\rm H\beta}$\tablenotemark{c}} &
    \colhead{$f_{\rm [OIII],5007}$\tablenotemark{c}} &
    \colhead{$f_{\rm [NII],6583}$\tablenotemark{c}} \\
    \colhead{} & \colhead{(mag)} & \colhead{(mag)} & \colhead{(arcsec)} & \colhead{(arcsec)} & \colhead{(deg)} & \colhead{(km\:s$^{-1}$)}
    }
	\startdata
    BG & 0.85 & 17.15 & 0.67 & 0.37 & \ldots & \ldots & $140.1\pm15.4$ & $45.8\pm4.2$ & $76.9\pm4.6$ & $15.2\pm3.5$ \\
	FG & 0.25 & 18.44 & 2.98 & \ldots& $44.2\pm1.7$ & 310 & $186.1\pm11.3$ & $31.7\pm6.8$ & $9.8\pm6.0$ & $62.1\pm9.0$ \\
	CG & 0.51 & 20.22 & 1.09 & \ldots & \ldots & \ldots & \ldots & \ldots & \ldots & \ldots
	\enddata
	\tablenotetext{a}{Half-light radius and FG inclination are derived from our MMT $r$-band image using GALFIT \citep{galfit1,galfit2}.}
	\tablenotetext{b}{FUV half-light radius \citep{alex15}.}
    \tablenotetext{c}{Line fluxes are in units of $\times 10^{-17}$ erg~s$^{-1}$~cm$^{-2}$ and are taken from the MPA-JHU DR7 catalog.
    }
\end{deluxetable*}

\subsection{Size of the DLA Cloud}

An extended background source like BG allows us to investigate the spatial scale of the DLA. Using a deep near-UV acquisition image obtained with the COS, \citeauthor{alex15} found the half-light radius of BG at 2433~$\rm \AA$ to be 0.37$^{\prime\prime}$ which at the redshift of the DLA corresponds to 1.09~kpc. We calculate a residual flux over the base of the \Lya profile ($-$334 to 594\:km\:s$^{-1}$) of $0.021 \pm 0.096$ (normalized). Using a 1$\sigma$ upper limit for the residual flux of 0.117 we find a lower limit on the covering fraction of 0.883 for the DLA. Multiplying this value by the projected area of BG in the DLA plane results in a lower limit for the size (area) of the DLA of $A_{DLA}\rm \geq3.3~kpc^2$.
Combined with the most conservative estimate for the \HI column density of $N($\ion{H}{1})$= 2.00 \times 10^{20}$~cm$^{-2}$ we estimate the mass of the DLA to be $M_{DLA}\geq5.3\times 10^6~M_\odot$ after correcting for helium.

Our estimates on the size and mass of the DLA are consistent with results from two high-redshift DLAs studied by \cite{cooke15} and \cite{mawa16} using Lyman-break galaxy (LBG) sightlines. With the uncertainty in the area of their background sources, \cite{cooke15} estimate a neutral gas mass of $\sim 10^6-10^9 M_\odot$ for their DLA at $\rm z\sim2.4$. \cite{mawa16} find a lower limit on the area of $\gtrsim1 \rm kpc^2$ for a DLA at $\rm z=3.335$. Though these DLAs are found at significantly higher redshifts, they are being detected using very similar background systems to BG. BG matches many properties of LBGs such as morphology, size, UV luminosity, and SFR \citep[][and references therein]{over09, alex15}. Hence, the match between our estimate of the limiting size of the DLA and those from higher redshift studies is not surprising. 

\subsection{DLA Environment} \label{sec:origin}
Our sightline passes through FG at an impact parameter of $\rho=36$\:kpc, which is at 10\% of the virial radius. 
This suggests that the DLA could be associated with (1) the \ion{H}{1} disk of FG, (2) condensing clouds in the CGM of FG, (3) a tidal structure generated by a recent interaction, or (4) high velocity outflows from BG. Here, we investigate these four possibilities.

\subsubsection{Extended \HI Disk}

First, let us investigate if the DLA is associated with the disk of FG. This would imply that the \HI disk of FG has a radius of at least 36~kpc with an average column density of $N($\ion{H}{1})$= 3.47 \rm \times 10^{20}~cm^{-2}$. By applying the tight correlation between \HI mass and \HI disk size \citep{broeils_rhee97, swaters02, wang16}, we can estimate the \HI mass of FG to be $M($\ion{H}{1}$)\rm > 1.5\times 10^{10}~M_{\odot}$. This would imply the gas fraction of FG is $\mathrm{log}~ M($\ion{H}{1}$)/M_{\star} = -0.92$, a value consistent with galaxies of similar mass at low redshifts in the GASS survey \citep{catinella10} and is comparable to the Milky Way.

The measured rotation of FG (Figure \ref{fig:rotation}c) also supports the scenario where the DLA is associated with the disk of the galaxy. We find the mean velocity of the flat portion of the projected disk rotation to be $v_f\approx \rm 215~km~s^{-1}$. The relative velocity of the DLA absorption is consistent with this projected velocity as the dominant component of the DLA is found at 131~km~s$^{-1}$ from the systemic of FG (blue star in Figure \ref{fig:rotation}c with the blue shaded region indicating $\pm90\%$ of the \ion{Si}{2} $\lambda1260$ profile width). The observed velocity offset may be explained by a drop in the rotation curve at this large distance (36 kpc) or warps in the extended disk. Rotation velocities are known to be lower in the extended disk at large radii \citep[e.g.][]{battaglia06}. Additionally, warps in \ion{H}{1} disks that are very commonly seen even in isolated galaxies and are known to exist well beyond the optical extent of the galaxy  \citep[e.g.][]{sancisi76,briggs90,binney92,garcia02}. A drop in velocity by $\sim$30 km~s$^{-1}$ and a change in inclination between the optical and \ion{H}{1} disk of 10$-$15 degrees can explain the observed velocity offset between the DLA and the optical rotation.

In a similar study as presented here, \cite{stanic16} used an extended background source to study \ion{Mg}{2} absorption at $\rho=27~\textrm{kpc}$ from a Milky Way analog galaxy at $z=0.4$ and found optically thick absorption with an estimated $N($\ion{H}{1})$\geq 10^{19}~\rm cm^{-2}$ that kinematically matches the galaxy rotation and is consistent with an extended or warped disk. Similar conclusions have also been made in QSO absorption line work. Using studies of L$^*$ galaxies at $z\leq0.2$ \cite{bregman18} finds a distinct population of $N$(\ion{H}{1}) absorbers that lie within 50~kpc of galaxies that are consistent with extended disks. Co-rotation is also seen in \ion{Mg}{2} absorption line studies \citep{ho17} and hydrodynamic simulations using EAGLE \citep{ho19} that show sightlines that pass close to the major axis of $z\leq0.3$ galaxies are likely tracing cool gas that is within an extended disk. Additionally, \cite{neeleman17} find a $z=4.254$ DLA at an impact parameter of 42 kpc from a star forming galaxy that is co-rotating with the disk.

The long-slit spectrum also shows evidence for recent star formation outside the main stellar disk as there is \Ha emission that is beyond the main component of the galaxy (black arrows in Figure \ref{fig:rotation}, a and b). The symmetry of this emission suggests a possible ring structure around FG at a distance of 18~kpc. The presence of the DLA could therefore be associated with feeding star formation. High column density gas related to star formation is not uncommon at large distances as studies of extended ultraviolet disks have shown star formation at radii up to five times larger than the extent of the optical disk \citep{thilker05, depaz05}.

The most likely scenario is that we are seeing the \HI disk of FG in absorption due to the kinematic similarities between the stellar disk and the DLA cloud, as well as the detected DLA species. The Doppler b-values, and the absence of high-ionization transitions like \ion{Si}{4} (ionization potential of $>45.1~eV$ required to produce \ion{Si}{4}), despite the presence of \ion{Si}{3}, would indicate that the gas is at low-to-intermediate ionization states and is most likely tracing the warm neutral medium. 
This conclusion is corroborated by studies that confirm the lack of \HI 21~cm absorption tracing cold neutral medium associated with galaxy disks beyond 15-20~kpc for most low-z galaxies  \citep{borthakur14, borthakur16, dutta17, dutta19}.

\subsubsection{Tidal Interactions}

The second possibility is that we might be detecting tidal debris resulting from galaxy interactions.
Interactions are known to play an important role in moving high column density ISM to large distances outside galaxies at low-z \citep{hibbard00}.
\cite{zwaan08} argued that DLAs are likely the result of tidal interactions or outflows and unlikely to be rotation in cold disks based on their comparison of DLA velocity profiles with those of \ion{H}{1} gas in galaxy disks.
Our VATT and MMTO imaging shows a stellar extension to the south of FG at faint surface brightness levels that could be a tidal structure as well as a clump of stellar emission nearly opposite to this extension to the northeast of the galaxy.
These regions could be due to tidal interactions with an unseen companion, or the galaxy seen to the southwest of BG, located a projected $\rho=66$\:kpc from FG. A photometric redshift of $z_{\mathrm{phot}}=0.19\pm0.07$ places it near FG however no spectroscopic redshift is available for this galaxy. If these two galaxies are indeed interacting, then we expect to find a tidal bridge connecting the two galaxies \citep{toom72} that would pass through our sightline where the DLA is detected.

One example of a DLA tracing tidal debris was presented by \citet{kacp10} in their survey of a QSO field with the \textit{HST} Wide Field Planetary Camera 2 (WFPC-2) of a known DLA system. They found multiple galaxies that showed perturbed morphologies and tidal tails that extended up to 25 kpc leading them to conclude that the DLA was tracing tidal debris.
More recently, \cite{augustin18} surveyed the fields of DLAs with the Wide Field Camera 3 aboard {\em HST} to investigate the morphological properties of the DLA host galaxies and found them to show clumpy structure suggestive of ongoing tidal interaction. 
Similar conclusions were also drawn by \citet{chen19} in their discovery of a spatially extended line-emitting nebula around a DLA that is associated with a galaxy group, and by \cite{borth19} that found a DLA at 65 kpc without detecting \ion{Si}{4}, resembling the present case.
There is no evidence that FG is part of a larger group, but there could be smaller/fainter galaxies near FG including the third galaxy in the field (southwest of FG and BG) for which we have neither deep imaging nor spectroscopic data.

\subsubsection{CGM Gas}

The third possibility is that the DLA is tracing CGM clouds. Studies of neutral \ion{H}{1} in the CGM of low-z galaxies have shown that absorbing gas is typically located $\rm \pm200 km~s^{-1}$ from the host galaxy systemic velocity \citep{tum13, liang2014, borth15}. The observed velocity difference of 131\:km\:s$^{-1}$ is well within the values seen in most of these surveys. However, it is worth noting that our sightline passes quite close to the galaxy, unlike in most \Lya studies.
The size and the mass of the DLA cloud is consistent with high- and intermediate-velocity clouds (HVCs, IVCs) seen in Milky Way, although the covering fraction of HVCs at even sub-DLA column densities is quite low \citep{wakker01}. Recent study of the column density distribution function of HVCs and IVCs by  \citet{french20} find that there are no absorbers at column densities greater than 10$\rm ^{20}~cm^{-2}$ while also finding that column densities above this limit is routinely seen in the ISM of the Milky Way. Therefore, assuming BG is similar to the Milky Way, it is likely that the DLA is part of the ISM of BG.

Studies of \ion{Mg}{2} absorption line systems detected in QSO spectra passing close in projection to foreground star-forming galaxies have found CGM absorption velocities are nearly always aligned with the rotation of the stellar disks \citep{steidel02, kacp10mg2, martin19}. Simple disk models cannot explain the observed velocity widths in many of these sightlines even though the majority are at low impact parameters with $\rho<0.5~R_{vir}$ \citep{kacp10mg2}. These studies are tracing metal-enriched gas, however, so are sensitive to outflowing or recycled material. Additionally, the strongest \ion{Mg}{2} absorbers detected require multiple velocity components rather than high column density gas like the DLA presented here \citep{martin19}. Although the properties of our system show similarities to these \ion{Mg}{2} studies we believe that it is unlikely we are observing a CGM cloud unrelated to the \ion{H}{1} disk as it would require the cloud to be coincidentally at the correct rotation velocity of FG or that we are detecting material that is actively accreting onto the disk.

\subsubsection{High Velocity Outflows from BG}
Here we discuss the evidence that led us to conclude that the DLA is not high-velocity outflowing gas from the background galaxy.
While the DLA velocity offset from BG of $\sim$1200\:km\:s$^{-1}$ is within the possible range seen in some of the rare high velocity outflows from starburst galaxies \citep[up to 2400~km~s$^{-1}$; ][]{tremonti09, rubin11, rubin14, chisholm15, heckman16}, the strength and kinematics of the DLA suggest that it is most likely the disk of FG.
Down-the-barrel studies showing outflows are typically dominated by broad absorption features as they are a convolution of multiple components at different velocities. Conversely, for our system the DLA absorption features are distinct from stellar absorption from BG (see the marked \ion{Si}{3} absorption in Figure \ref{fig:dla}). Additionally, the plasma from galactic outflows is typically multiphase showing gas at multiple different ionization states \citep{tremonti09, alex15, heckman16}. As we do not detect \ion{Si}{4} absorption despite having high column components of \ion{Si}{2} and \ion{Si}{3}, it is unlikely we are tracing outflows from BG as it would require an anomalous cloud of cooler gas that is being ejected from the galaxy at high speeds. The detected species and distinct velocity offset suggest we are tracing cool gas that is most likely  entrained in the disk of FG. In addition, the maximum outflow velocity of BG was estimated to be 510~\kms \citep{heckman16}, which is consistent with the relationship between outflow velocity and the star-formation rate surface density of the host galaxy. Therefore, based on the properties of BG and those of the DLA, we conclude that the DLA is not associated with high-velocity outflows.

\section{Conclusions} \label{sec:conclusion}
We have presented the results of a study of a low-redshift DLA found using a starburst galaxy background source. This is the first such detection at low-redshifts among three ever detected \citep{cooke15, mawa16} and the first with a confirmed host galaxy. The background galaxy (BG) sightline is located a projected 36\:kpc from the DLA host galaxy, FG, at $\rm z=0.17077$. FG is a star forming galaxy with a specific star formation rate of $\rm log~sSFR=-9.15$ and does not show any AGN signatures. We find that:

\begin{enumerate}
	\item The detected DLA has a hydrogen column density of $\mathrm{log}N($\ion{H}{1}$)=20.5^{+0.2}_{-0.2}$. We also detect  metal-line transitions of \ion{N}{1}, \ion{N}{2}, \ion{Si}{2}, \ion{C}{2}, and \ion{Si}{3}. With the exception of \ion{N}{1}, each transition contains two components. Additionally, we do not detect \ion{N}{5}, \ion{Si}{4}, \ion{Ca}{2}, or \ion{Na}{1} suggesting the gas is at low-to-intermediate ionization states. 
	\item Using the covering fraction of the DLA and the half-light radius of BG, the DLA is found to have an area of $A_{DLA}>3.3~\rm kpc^2$ and a neutral hydrogen mass of $M_{DLA}>5.3\times 10^6~M_\odot$. Both these values are consistent with the two previously detected DLAs using extended background sources.
	\item The DLA is located +131\:km\:s$^{-1}$ from the systemic velocity of FG and the absorption velocity is aligned with the rotation velocity of the disk closest to the sightline ($v_f\approx \rm 215~km~s^{-1}$). It is therefore likely that we are observing the disk of FG. The measured neutral hydrogen column density, gas kinematics, and the impact parameter of the QSO sightline of 36 kpc (10\% $R_{\rm vir}$) is consistent with gas in an extended \HI disk.
\end{enumerate}

 Using spatially extended background sources will become more prevalent as we look towards the next generation of telescopes and instruments as we will have a greater ability to obtain high quality spectra of faint sources. This is especially true at high redshifts ($z>2$) where the space density of QSOs decreases while the density of starbursting LBGs increases. Large spectroscopic surveys conducted with 30-meter class telescopes will enable us to study galaxies with large rest-UV flux that will have the potential to discover countless new DLAs. With the increased number of sightlines and the ability to probe the spatial extent of absorbers, we will be able to study the evolution of neutral gas clouds with unprecedented detail over a large redshift range using tracers such as \ion{Na}{1}~D, \ion{Ca}{2}, \ion{Mg}{2}, and \Lya.

\acknowledgments
We thank the referee for their constructive comments. We would like to thank the staff at VATT and MMTO for their help during preparations for our observations. We thank members of the STARs lab (ASU) for their feedback during discussions of this project. The Arizona State University authors acknowledge the twenty-two Native Nations that have inhabited this land for centuries. Arizona State University's four campuses are located in the Salt River Valley on ancestral territories of Indigenous peoples, including the Akimel O’odham (Pima) and Pee Posh (Maricopa) Indian Communities, whose care and keeping of these lands allows us to be here today. We acknowledge the sovereignty of these nations and seek to foster an environment of success and possibility for Native American students and patrons. 

This work is based on observations with the NASA/ESA Hubble Space Telescope, which is operated by the Association of Universities for Research in Astronomy, Inc., under NASA contract NAS5-26555. Some of the observations reported here were obtained at the MMT Observatory, a joint facility of the Smithsonian Institution and the University of Arizona. This project is also based in part on observations with the VATT: the Alice P. Lennon Telescope and the Thomas J. Bannan Astrophysics Facility. This project additionally made use of SDSS data. Funding for the Sloan Digital Sky Survey IV has been provided by the Alfred P. Sloan Foundation, the U.S. Department of Energy Office of Science, and the Participating Institutions. SDSS-IV acknowledges
support and resources from the Center for High-Performance Computing at
the University of Utah. The SDSS web site is www.sdss.org.
SDSS-IV is managed by the Astrophysical Research Consortium for the 
Participating Institutions of the SDSS Collaboration including the 
Brazilian Participation Group, the Carnegie Institution for Science, 
Carnegie Mellon University, the Chilean Participation Group, the French Participation Group, Harvard-Smithsonian Center for Astrophysics, 
Instituto de Astrof\'isica de Canarias, The Johns Hopkins University, Kavli Institute for the Physics and Mathematics of the Universe (IPMU) / 
University of Tokyo, the Korean Participation Group, Lawrence Berkeley National Laboratory, 
Leibniz Institut f\"ur Astrophysik Potsdam (AIP),  
Max-Planck-Institut f\"ur Astronomie (MPIA Heidelberg), 
Max-Planck-Institut f\"ur Astrophysik (MPA Garching), 
Max-Planck-Institut f\"ur Extraterrestrische Physik (MPE), 
National Astronomical Observatories of China, New Mexico State University, 
New York University, University of Notre Dame, 
Observat\'ario Nacional / MCTI, The Ohio State University, 
Pennsylvania State University, Shanghai Astronomical Observatory, 
United Kingdom Participation Group,
Universidad Nacional Aut\'onoma de M\'exico, University of Arizona, 
University of Colorado Boulder, University of Oxford, University of Portsmouth, 
University of Utah, University of Virginia, University of Washington, University of Wisconsin, 
Vanderbilt University, and Yale University.

\pagebreak
\bibliography{J1113_apj}{}
\bibliographystyle{aasjournal}

\end{document}